\documentclass[aip,rsi,amsmath,amssymb,longbibliography,preprint]{revtex4-1} 
\usepackage{bm}
\usepackage{graphicx}
\usepackage{algorithm}
\usepackage{float}
\usepackage{algpseudocode}
\usepackage{verbatim}
\usepackage{subcaption}
\usepackage{color}
\usepackage{hyperref}
\usepackage{layouts}

\def\e{\mathrm{e}}

\begin{document}



\title[Expectation--maximization algorithm for PEPT]{An expectation--maximization algorithm for positron emission particle tracking}

\author{Sam Manger}
\email{samuelpmanger@gmail.com}
\affiliation{School of Chemical Engineering, University of Birmingham, B15 2TT, UK}

\author {Antoine Renaud}
\affiliation{School  of  Mathematics  and  Maxwell  Institute  for  Mathematical  Sciences,  University  of  Edinburgh, EH9 3FD, UK}

\author{Jacques Vanneste}
\affiliation{School  of  Mathematics  and  Maxwell  Institute  for  Mathematical  Sciences,  University  of  Edinburgh, EH9 3FD, UK}

\date{\today}

\begin{abstract}
Positron Emission Particle Tracking (PEPT) is an imaging method that tracks individual radioactive particles. PEPT relies on the detection of back-to-back photon pairs emitted by positron annihilation. It requires an algorithm to locate the radioactive particles based on the set of lines defined by successive photon-pair detections. We propose and test a new algorithm for this task.

The algorithm relies on the maximization of a likelihood arising from a simple Gaussian-mixture model defined in the space of lines. The model includes a component that accounts for spurious lines caused by scattering and random coincidence, and treats the relative activity of particles as well as their positions as parameters to be inferred. Values of these parameters that approximately maximize the likelihood are computed by application of an expectation-maximization algorithm. A generalization of the model that includes the particle velocities and accelerations as additional parameters takes advantage of the information contained in the exact timing of positron annihilations to reconstruct pieces of trajectories rather than fixed positions, with clear benefits.

We test the algorithm on both simulated and experimental data. The results show the algorithm to be highly effective for the simultaneous tracking of many particles (up to 80 in one test). It provides estimates of particle positions that are easily mapped to entire trajectories and handles a variable number of particles in the field of view. The ability to track a large number of particles robustly offers the possibility of a dramatic expansion of the scope of PEPT

\end{abstract}


\maketitle

\section{Introduction}
Positron Emission Particle Tracking (PEPT) is a method of tracking single particles using the back-to-back emission of 511 keV photons from positron annihilation. The PEPT technique uses similar principles to Positron Emission Tomography (PET), in which a radioactive tracer undergoing $\beta^+$ decay emits positrons which subsequently annihilate with surrounding electrons, generating the back-to-back photons.  The detection of these gamma ray pairs produces data in the form of  lines in $\mathbb{R}^3$, known as Lines of Response (LORs). These LORs can then be used to determine the location of the radioactive tracer material. The highly penetrating nature of the 511 keV photons makes PEPT and PET ideal techniques for studying the internal dynamics of opaque systems, where optical techniques fail. However, there is a non-negligible probability that one photon may have undergone a Compton scattering interaction with surrounding material. Alternatively, a random coincidence may be detected, where the instrumentation attributes a LOR to two uncorrelated photons. These scattered and random LORs may not cross the source, so complicate the tracking process.

Typically, the PET technique consists of inverting the detected LORs to produce a three-dimensional image of radioactive scalar density by attributing levels of radioactivity to each voxel in the field of view. This requires a relatively long exposure time and is only suitable for visualising dynamic processes occurring on time scales of the order of 1 second. In contrast, in PEPT the radioactive sources are typically small particles less than a millimetre in size that can be treated as point-like sources. This strong prior allows us to dramatically reduce the amount of data required, and thus the exposure time, to accurately locate the particles. This makes it possible to track particle movement with significantly higher time resolution, to the order of 1 millisecond. This makes PEPT a powerful technique to study the dynamics of particles in fluid flows, especially particle-laden flows, molten metals or flows in opaque containers
for which optical methods are unavailable\cite{Barigou2004,Fangary2002}.

Locating a single particle in PEPT consists of clustering the line data into an  \emph{inlier} and an \emph{outlier} set, in which the \emph{inlier} set contains the ``true'' LORs and the \emph{outlier} set contains spurious lines generated by scattered photons or random coincidences. The particle position is then the centroid of the \emph{inlier} set. The pioneering and widely used algorithm developed by David Parker et al \cite{Parker1993} at the University of Birmingham performs such a clustering by iteratively trimming the \emph{inlier} cluster from its lines furthest from its centroid. However, to localize multiple particles at once several \emph{inlier} clusters must be found. A generalization of the Birmingham algorithm has been developed to localize multiple particles sequentially but it fails to localize more than just a few particles. Other heuristic approaches have been developed, using e.g., feature voxel detection \cite{Bickell2012}, Voronoi tessellations \cite{Blakemore2019}, G-means clustering \cite{wiggins2016}, feature point identification \cite{wiggins2017}, spherical density analysis \cite{Odo2019} and machine learning \cite{Nicusan2020}. Ref.\ \citenum{Nicusan2020} gives a convenient review of these approaches before developing an alternative based on cut-point clustering.

In this paper, we develop a new algorithm to infer the position of multiple PEPT particles from detected LORs. The new algorithm departs from earlier ones in three significant ways. First, the clustering is carried out in the space of LORs rather than a space of representative points such as cut points (closest points between pairs of lines) or high-intensity voxels. Second, the method is anchored in Bayesian inference and based on a model of the physics of positron emission -- as a result, the activity of each particle is inferred along with its coordinates. Third, the information provided by timestamp of detected LORs can be integrated into the model to infer particle velocities and accelerations in addition to positions, thus providing a direct approximation to the trajectories of the particles over short times.  
At its core, the algorithm relies on the maximization of the likelihood of a Gaussian mixture model. This is carried out using the expectation--maximization (EM) algorithm, a soft-clustering method well adapted to problems of this type \cite{EM_orgin,EM_Review,bilmes1998,mclachlan2007}.

%
The paper is organized as follows. In section \ref{sec:GMM}, we present the Gaussian mixture model we propose to describe the generation of LORs by the radioactive particles, incorporating a crude representation of scattering. In section \ref{sec:Localizing}, we introduce the new algorithm, which we term PEPT-EM, and demonstrate its capability on a basic example. In section \ref{sec:Tracking}, we generalize the algorithm to track multiple moving particles, incorporating temporal information via velocity and acceleration inference, and  we apply it to simple cases. In section \ref{sec:Application}, we test the PEPT-EM algorithm on simulated and experimental PEPT data. We conclude in section \ref{sec:Conclusion}.

\section{Mixture model\label{sec:GMM}}

We model the generation of LORs as a result of positron emission as follows. Suppose there are $K$ radioactive particles located at positions $\bm{x}_k, \, k=1,\cdots,K$ and assume that, when a particle emits a positron, this is annihilated at a nearby point $\bm{y}_k$ distributed according to a Gaussian distribution centered at $\bm{x}_k$ with variance $\sigma_k^2$. The probability that the annihilation occurs at a point $\bm{y}$ can then be written as 
\begin{equation}\label{eq:GM_Rd}
    P\left(\bm{y}\left|\left\{\bm{x}_k,\sigma^2_k,\rho_k\right\}_{1\leq k\leq K}\right.\right)=\sum_{k=1}^K\frac{\rho_k}{(2\pi\sigma_k^2)^{3/2}} \, \mathrm{e}^{-{(\bm{y}-\bm{x}_k)^2}/2\sigma_k^2} \,,
\end{equation}
where $\rho_k$ is proportional to the activity of particle $k$, with $\sum_{k=1}^{K}\rho_k=1$. Eq.\ \eqref{eq:GM_Rd} is the probability density of a Gaussian mixture, with $\rho_k$ interpreted as the weight of component $k$ of the mixture. 



\begin{figure}
    \centering \includegraphics[width=0.4\linewidth]{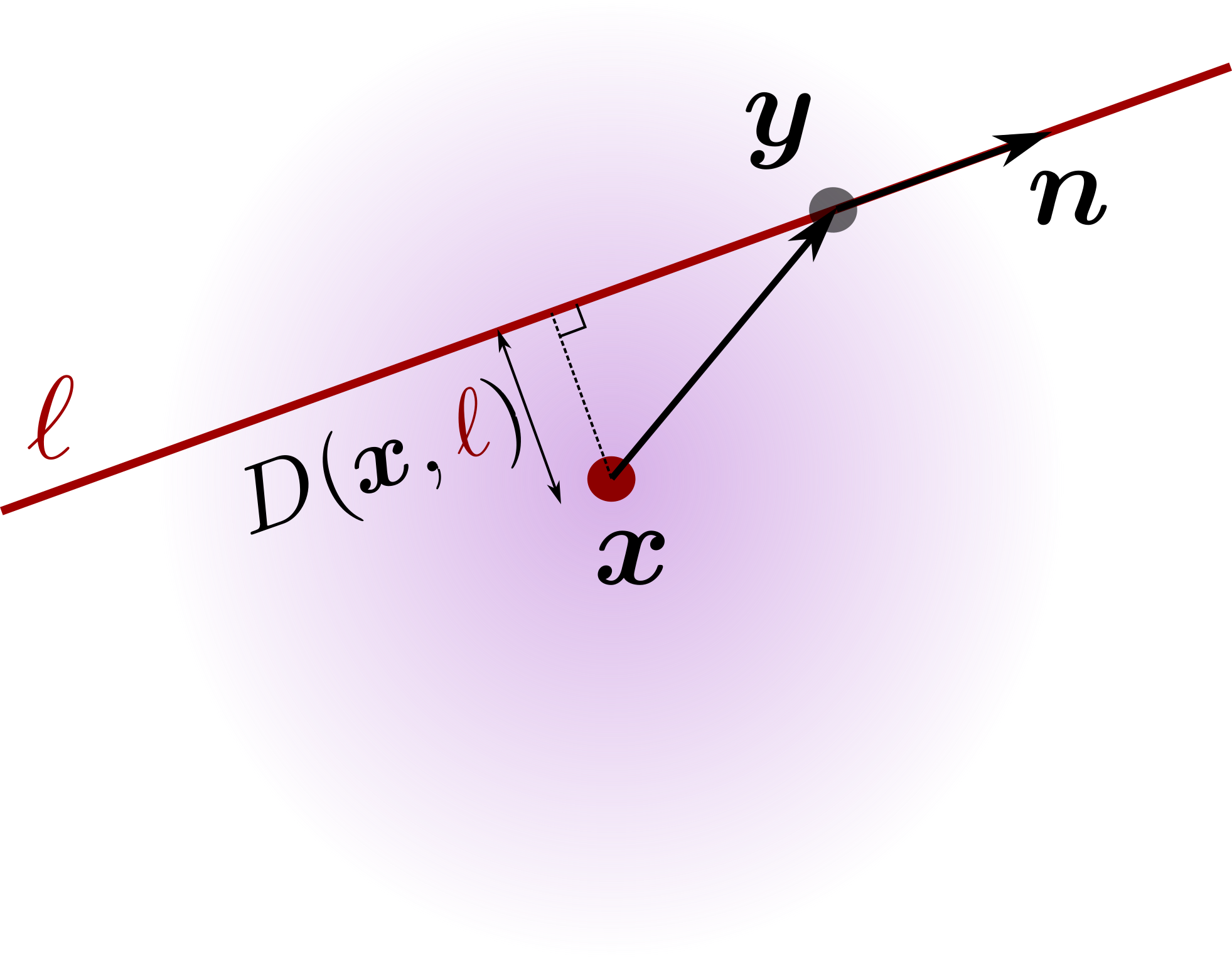}
    \caption{Schematic of the generation of a LOR: a radioactive particle located at $\bm{x}$ emits a positron which is annihilated at a point $\bm{y}$ drawn from a Gaussian distribution centered at $\bm{x}$; the direction $\bm{n}$ of the LOR is distributed uniformly.}
    \label{fig:FwMod_Sketch}
\end{figure}

A positron annihilation at $\bm{y}$ produces back-to-back photons travelling in a random direction. We represent this direction by a unit vector $\bm{n}$  distributed uniformly over the unit hemisphere.
The associated LOR is then described by the parametric equation $\bm{x} = \bm{y} + \lambda \bm{n}$, with $\bm{x}$ an arbitrary point on the line and $\lambda \in \mathbb{R}$ the parameter. We characterize the LOR by the two vectors $\bm{y}$ and $\bm{n}$, and write this as $\ell=\left|\bm{y},\bm{n}\right)$. Observe that the characterization is not unique because $\ell=\left|\bm{y}+\lambda\bm{n},\bm{n}\right)$ for any $\lambda\in\mathbb{R}$ corresponds to the same line as $\left|\bm{y},\bm{n}\right)$. The probability of a particular LOR $\ell$ is then obtained by integrating \eqref{eq:GM_Rd} over all annihilation points which potentially generated this LOR, leading to
\begin{equation}\label{eq:ProbLine_yn}
    P\left(\ell\left|\left\{\bm{x}_k,\sigma^2_k,\rho_k\right\}_{1\leq k\leq K}\right.\right)= \frac{1}{2\pi} \int_{\mathbb{R}}\mathrm{d}\lambda\,P\left(\bm{y}+\lambda\bm{n}\left|\left\{\bm{x}_k,\sigma^2_k,\rho_k\right\}_{1\leq k\leq K}\right.\right),
\end{equation}
where the factor $1/(2\pi)$ accounts for the uniform distribution of $\bm{n}$ over a unit hemisphere. Evaluating the integral in \eqref{eq:ProbLine_yn} gives
\begin{equation}\label{eq:ProbLine}
    P\left(\ell\left|\left\{\bm{x}_k,\sigma^2_k,\rho_k\right\}_{1\leq k\leq K}\right.\right)=\sum_{k=1}^K \frac{\rho_k}{4\pi^2\sigma_k^2} \, \mathrm{e}^{-D^2(\bm{x}_k,\ell)/2\sigma_k^2},
\end{equation}
where 
    \begin{equation}\label{eq:D2x}
        D^2(\bm{x},\ell)=\left|\bm{x}-\bm{y}\right|^2-\left(\left(\bm{x}-\bm{y}\right)\cdot\bm{n}\right)^2
    \end{equation}
is the shortest square distance between the point $\bm{x}$ and the line $\ell$, which is independent of the specific characterization $\ell=|\bm{y},\bm{n})$ used. The distribution \eqref{eq:ProbLine} can be interpreted as a Gaussian mixture model in the space of lines.
Figure \ref{fig:FwMod_Sketch} illustrates this model for a single component.

We account for outliers, that is, spurious LORs which result from Compton scattering or random coincidences, in a simple way. We add a component $k=0$ to the mixture model \eqref{eq:ProbLine}, with $\rho_0$ denoting the probability that a LOR belongs to this component, in other words that it
be an outlier, and we assume that outliers are distributed uniformly in the space of observable LORs. The probability of a LOR becomes
\begin{equation}\label{eq:LORProb}
P\left(\ell \left|\left\{\bm{x}_k,\sigma^2_k,\rho_k\right\} _{1\leq k\leq K}\right.\right) \propto \rho_0\alpha+\sum_{k=1}^{K}\rho_k\sigma_k^{-2}\mathrm{e}^{-{D^2(\bm{x}_k,\ell)}/2\sigma_k^2},
\end{equation}
ignoring an irrelevant constant factor. Here $\alpha$ is a constant fixed \emph{a priori} and interpreted as the inverse variance of the outlier cluster. In practice, $\alpha^{-3/2}$ is of the order of the volume of the field of view. 
The probability $\rho_0$ is determined from the other $\rho_k$ via the normalization condition $\sum_{k=0}^K \rho_k = 1$.

The probability of observing a set of lines $\mathcal{L}=\{\ell_1,\cdots,\ell_N \}$ is simply the product of $N$ factors of the form \eqref{eq:LORProb}. Taking the logarithm leads to the log-likelihood function
\begin{equation}\label{eq:LogLH_alpha}
   L\left(\left. \{\bm{x}_k,\sigma_k^2,\rho_k\}_{1\leq k\leq K} \right| \mathcal{L} \right) 
   =\sum_{\ell\in\mathcal{L}}\log\left(\rho_0\alpha+\sum_{k=1}^{K}\rho_k\sigma_k^{-2}\mathrm{e}^{-{D^2(\bm{x}_k,\ell)}/2\sigma_k^2}\right).
\end{equation}
In principle, maximizing this function with respect to its $5K$ arguments gives the most likely positions $\bm{x}_k$ of the radioactive particles and most likely weights $\rho_k$, proportional to the particles' activities. In a Bayesian framework the contribution of a prior should be added but we do not introduce one, effectively assuming it to be uniform. The global maximization of \eqref{eq:LogLH_alpha} for more than a few particles is an extremely challenging numerical problem  because of the high-dimensionality and presence of numerous local maxima. Instead we employ the EM algorithm which maximizes a relaxed version of \eqref{eq:LogLH_alpha} and typically provides a good approximation to the true maximizer. We describe and test the algorithm next.

 \section{Expectation--maximization \label{sec:Localizing}}


\subsection{Algorithm} 

The EM algorithm \cite{EM_orgin,EM_Review,bilmes1998,mclachlan2007} is tailored to mixture models described by a log-likelihood such as \eqref{eq:LogLH_alpha}. It introduces unobserved (latent) random variables $\zeta_\ell \in \{0,1,\cdots,K\}$  associated with each observed LOR $\ell$ and such that $\zeta_\ell = k$ if $\ell$  has been generated by the annihilation of a positron emitted by particle $k$ (or is an outlier if $k=0$). \(\zeta_\ell\) is a variable that gives the component of the Gaussian mixture model from which \(\ell\) is drawn. It serves as a kind of intermediate step to understand the EM algorithm. For fixed parameters $\{\bm{x}_k,\sigma^2_k,\rho_k\}_{1 \le k \le K}$, the probability that $\zeta_\ell = k$ is given by the so-called latent weights,
\begin{equation}\label{eq:Laten_Weights}
    w_{\ell,k}\left(\left\{\bm{x}_{k'},\sigma_{k'}^2,\rho_{k'}\right\}_{1\leq k'\leq K}\right)=\frac{ \rho_k\sigma^{-2}_k\mathrm{e}^{{-D^2(\bm{x}_k,\ell)}/2\sigma^2_k}}{ \rho_0\alpha+\sum_{k'=1}^K\rho_{k'}\sigma^{-2}_{k'}\mathrm{e}^{-{D^2(\bm{x}_{k'},\ell)}/2\sigma^2_{k'}}}
\end{equation}
for $1 \le k \le K$ and $w_{\ell,0} = 1 - \sum_{k=1}^K w_{\ell,k}$. Note that 
the log-likelihood \eqref{eq:LogLH_alpha}, when conditioned on the latent variables $\zeta_\ell$, takes a simple form:
    \begin{equation}\label{eq:condexp}
        L\left(\{\bm{x}_k,\sigma^2_k,\rho_k\}_{1 \le k \le K} |  \mathcal{L},\{\zeta_\ell\}_{\ell \in \mathcal{L}} \right) = \sum_{\ell \in \mathcal{L}} \left\{ \begin{array}{ll} \log \left( \rho_{\zeta_\ell} \sigma^{-2}_{\zeta_\ell} \e^{-D^2(\bm{x}_{\zeta_\ell},\ell) /2 \sigma^2_{\zeta_\ell}} \right) & \textrm{if } \zeta_l \ge 1 \\
        \log (\rho_0 \alpha)  & \textrm{if } \zeta_l = 0
        \end{array}
    \right. .
    \end{equation}
EM iterates over the values of the parameters $\{\bm{x}_k,\sigma^2_k,\rho_k\}_{1 \le k \le K}$, maximizing at each step the expectation of \eqref{eq:condexp} over the latent variables $\zeta_\ell$, assuming them distributed according to the latent weights \eqref{eq:Laten_Weights} obtained at the previous iterate. This expectation is given by
\begin{equation}
\label{eq:EME}
\mathbb{E}_{\{\zeta_\ell\}}\left[ L\left(\{\bm{x}_k,\sigma^2_k,\rho_k\}_{1 \le k \le K} \right. \right. |  \mathcal{L} \big) \big] \nonumber= \\\sum_{\ell \in \mathcal{L}}   \left( w_{\ell,0} \log (\rho_0 \alpha) + \sum_{k=1}^K w_{\ell,k} \log \left( \rho_k \sigma_k^2 \e^{-D^2(\bm{x}_k,\ell)/2 \sigma_k^2} \right) \right)
\end{equation}
and its maximization  can be carried out explicitly. The optimal positions $\bm{x}_k$ of the particles are found to be the weighted centroids of the LORs,
\begin{equation}\label{eq:Centroid_weighted}
           \bm{x}_k =  \mathbf{Centroid}_k \equiv \left(\sum_{\ell\in\mathcal{L}}w_{\ell,k}\mathbb{T}_{\ell}\right)^{-1}\sum_{\ell\in\mathcal{L}}w_{\ell,k}\mathbb{T}_{\ell}\,\bm{y}_{\ell},
           \\\quad
\text{with}\quad \mathbb{T}_{\ell}=\mathbb{I}_3-\bm{n}_{\ell}\otimes\bm{n}_{\ell}.
\end{equation}
 Here $\mathbb{I}_3$ is the identity matrix and $\bm{y}_{\ell}$ and $\bm{n}_{\ell}$ characterize the LOR $\ell$, that is, $\ell = |\bm{y},\bm{n} \rangle$. The optimal variances $\sigma_k^2$ are given by
        \begin{equation}\label{eq:Var_weigthed}
        \sigma^2_k = \mathrm{Var}_k \equiv \frac{\sum_{\ell\in\mathcal{L}}w_{\ell,k}D^{2}\left(\mathbf{Centroid}_k,\ell\right)}{2\sum_{\ell\in\mathcal{L}}w_{\ell,k}} .
        \end{equation}
Finally, the optimal component weights $\rho_k$ are given by 
        \begin{equation}\label{eq:R}
         \rho_k =   \mathrm{R}_k \equiv \frac{1}{N}\sum_{\ell\in L}w_{\ell,k},
        \end{equation}
with $N$ the number of LORs. The weight of the outliers is determined as      $\rho_0 = 1-\sum_{k=1}^K\mathrm{R}_k$.

With the explicit formulas \eqref{eq:Centroid_weighted}--\eqref{eq:R}, the EM  algorithm is straightforward. It starts by intializing the parameters  $\{\bm{x}_k,\sigma_k,\rho_k\}_{1\leq k\leq K}$. Then two steps are repeated alternatively until convergence: (i) the``expectation'' step computes the  latent weights $w_{\ell, k}$ from \eqref{eq:Laten_Weights}; and  (ii) the ``maximization'' step updates the parameters $\{\bm{x}_k,\sigma_k,\rho_k\}$ using \eqref{eq:Centroid_weighted}, \eqref{eq:Var_weigthed} and \eqref{eq:R} with the current  values of the $w_{\ell,k}$.

    \begin{algorithm}[H]
    \caption{PEPT-EM}\label{algo:EM_MM}
    \begin{algorithmic}[1]
     \State{Initialize $\{\bm{x}_k,\sigma^2_k,\rho_k\}_{1\leq k\leq K}$}
        \State{$\rho_0\gets1-\sum_{k=1}^K\rho_k$}
        
        \Repeat\Comment{Main loop}
        \State{compute $w_{k, \ell}$ from \eqref{eq:Laten_Weights}}\Comment{Expectation step}
            \For{$1\leq k\leq K$} \Comment{Maximization step}
             
            \If{Constrained} 
            \State{$\bm{x}_k,\sigma^2_k\gets\mathbf{Centroid}_k,\mathrm{Var}_k$}
            \Else
            \State{$\bm{x}_k,\sigma^2_k,\rho_k\gets\mathbf{Centroid}_k,\mathrm{Var}_k,\mathrm{R}_k$}
            \State{$\rho_{0}\gets1-\sum_{k=1}^{K}\rho_k$}
            \EndIf
            \EndFor
        \Until{convergence}
     \end{algorithmic}
\end{algorithm}

We note that, while we have so far identified $K$ with the number of particles $n_\mathrm{part}$, it is often advantageous to take $K > n_\mathrm{part}$ (and necessary when $n_\mathrm{part}$ is not known). We then interpret $K$ as a number of clusters, some of which capture outlier LORs. Particles are  associated only with clusters (i.e.\ mixture components) $k$ whose variance $\sigma_k^2$ are below a specified threshold, while the others  are interpreted as outliers, together with the $k=0$ component. The number of clusters $K$ and $\alpha$ are the only free parameters in the algorithm. These parameters can be tuned to reconstruct as many trajectories as possible while maintaining computational speed and convergence. The specific choice for $\alpha$ is less important when $K$ is sufficiently large. 
In practice, to facilitate the convergence of the algorithm, it is sometimes useful to constrain the weights $\{\rho_k\}_{0\leq k\leq K}$ to prescribed values. This constraint may be relaxed in the final algorithm steps. A pseudo-code implementation of our PEPT-EM algorithm including this possible constraint is given as Algorithm 1.

\subsection{Simple demonstration}
We first test the EM algorithm on synthetic LOR data  generated according to the Gaussian mixture model of section \ref{sec:GMM}. We use $n_{\rm part}$ particles drawn uniformly within a sphere
of radius $100$ centered at the origin, with identical standard deviations $\sigma_{k}=5$. An outlier component is added as an extra particle located at $\bm{x}_0=0$ with standard deviation $\sigma_0=100$ and weight $\rho_0=0.3$. The weights of the $n_{\rm part}$ are identical and given by $\rho_{k}=(1-\rho_0)/n_{\rm part}$. We draw
$N=n_{\rm part}\times 100$ lines from this model and apply the EM algorithm to recover the particle positions $\bm{x}_k$ from the LORs only.

  \begin{figure*}
    \centering
    \includegraphics[width=0.99\linewidth]{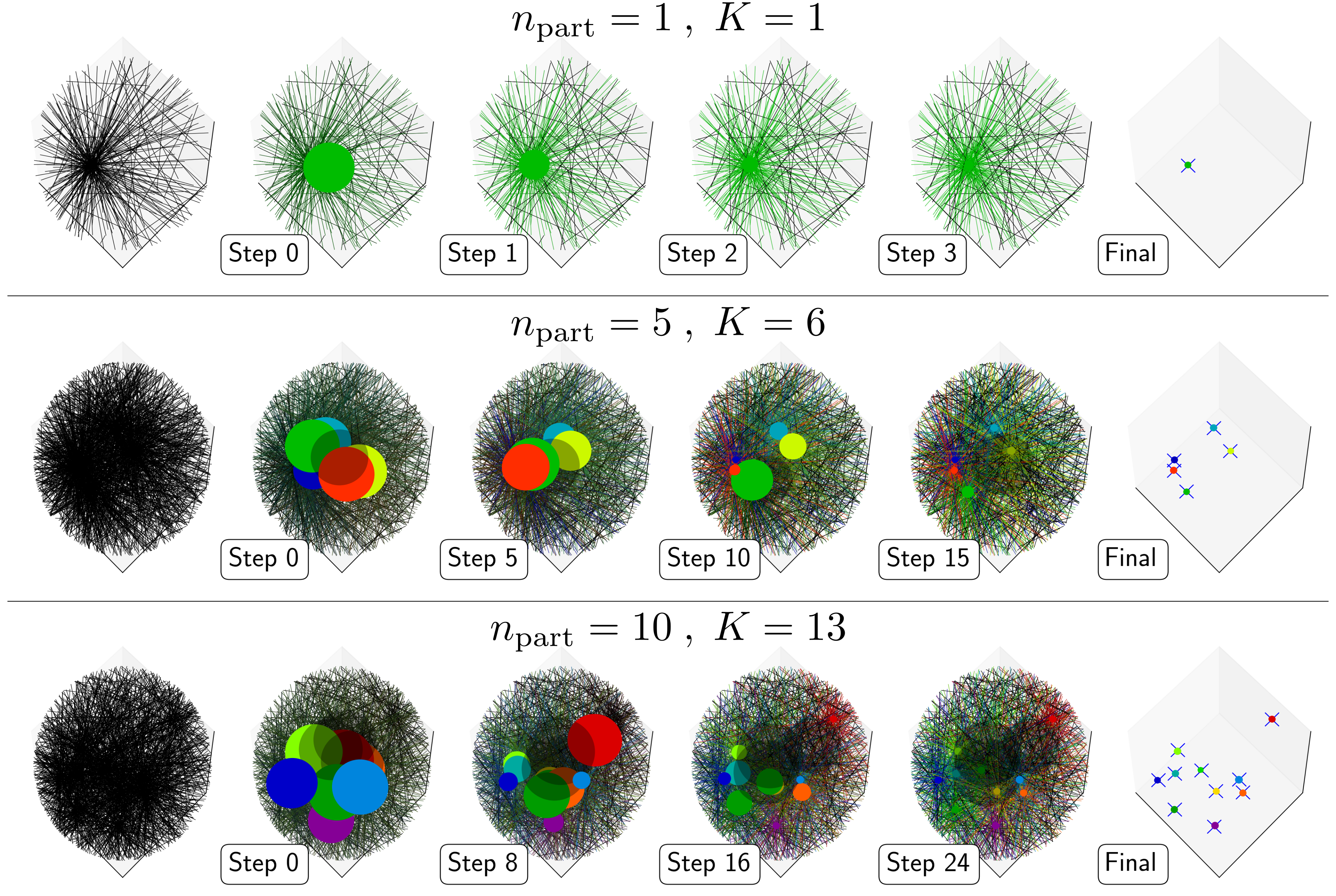}
    \caption{PEPT-EM algorithm applied to LORs  generated synthetically using the Gaussian mixture model of section \ref{sec:GMM}, with a number of particles $n_\textrm{part}=1$, $5$ and $10$ and one outlier component (see main text for details). The LORs are shown in the leftmost column. The rightmost column displays the particles' exact positions as blue crosses and the positions estimated by the PEPT-EM algorithm as colored spheres.  The middle columns show the locations $\bm{x}_k$ and standard deviations $\sigma_k$ after the indicated number of iterations as colored spheres of radius $\sigma_k$ centered at $\bm{x}_k$ (a mimimum value of the radius is set as $5$ for visibility). Colors are used for the $n_\textrm{part}$ mixture components with the smallest variances -- interpreted as particles -- and black for the remaining components -- interpreted as outliers. 
    The outliers are not shown in the rightmost column.}
    \label{fig:Illustration}
 \end{figure*}
 
The application of the PEPT-EM algorithm for different particle numbers $n_\textrm{part}$ and component numbers $K \ge n_\textrm{part}$ is illustrated in figure \ref{fig:Illustration}. The weight of the outlier component $k=0$ is $\alpha=10^{-4}$. We used the constrained algorithm with $\rho_k=1/(K+1)$ for $0\leq k\leq K$ and then applied the unconstrained algorithm.  The algorithm converges rapidly,  locating the particles to within an error of about $0.4$ which corresponds roughly to the standard deviation expected from the law of large number, $\sigma_{k}/\sqrt{N \rho_k}\approx0.4$. The figure illustrates how the variance of each component associated with a particle reduces as the iteration progresses, leaving high-variance clusters that capture outlier LORs.

\section{Particle tracking \label{sec:Tracking}}

Because PEPT is applied to particles that are moving, we need to reconstruct full time-dependent trajectories rather than fixed positions. A timestamp $t_\ell$ is associated with the observation of each LOR $\ell$. A space-time statistical model of the LORs, generalizing the model of section \ref{sec:GMM} to include the (exponentially distributed) times $t_\ell$, can be written down, but reconstructing entire trajectories on this basis, e.g.\ by attempting to maximize their likelihood, is a hopelessly complicated task. Instead, we adopt a sequential approach and infer a sequence of snapshots of the particle positions from which trajectories are reconstructed \textit{a posteriori}.



\subsection{Sequential tracking}

To infer the positions of the particles at regularly spaced times $t_n=n \delta t, \, n=1,2\cdots n$, with $\delta t$ the time interval, the observed LORs are grouped in small batches
\begin{equation}
    \mathcal{L}_{n}=\left\{\ell\in\mathcal{L}\left|t_\ell\in\left(n\delta t-{\Delta t}/{2},n\delta t+{\Delta t}/{2}\right)\right.\right\}.
\end{equation}
Here $\Delta t$ is an observation time that should be short enough that the particles can be considered as fixed over $\Delta t$, but long enough that the particle positions can be inferred accurately.
A straightforward approach simply applies the PEPT-EM algorithm to obtain the positions $\bm{x}_{k,n}$, variances $\sigma^2_{k,n}$ and densities $\rho_{k,n}$ for each batch $n$ which we associate with time $t_n$. This can done independently for each $n$, in parallel, and followed by a trajectory reconstruction procedure to map particles from one $t_n$ to the next.
However, a sequential treatment is preferable:
in most cases, the particles positions do not change much  between  successive times $t_n$, making the output of a batch a very good initial guess for the next batch; this considerably reduce the number of steps required for the PEPT-EM algorithm to converge. Another benefit of the sequential treatment is that it provides automatically a very good first guess for the trajectory reconstruction, since the cluster labels $k$ for successive batches generally correspond to the same particles.
In practice, moving particles can enter and exit the field of view, so that the number of particles $n_\mathrm{part}$ is unkown and can change between batches. 
This is handled by using a number of clusters $K$ larger than the typical number of particles and using only the clusters with variances $\sigma_k^2$ below a threshold to identify particle trajectories. When a particle leaves the field of view, the variance of the corresponding cluster increases and the association particle/cluster is broken. The cluster then becomes available for new particles as they enter the field of view. In this way, the sequential algorithm identifies segments of trajectories whose number can change between batches while keeping a fixed number of clusters $K$ throughout.


\begin{figure}
    \centering
        \includegraphics[height=0.45\textwidth]{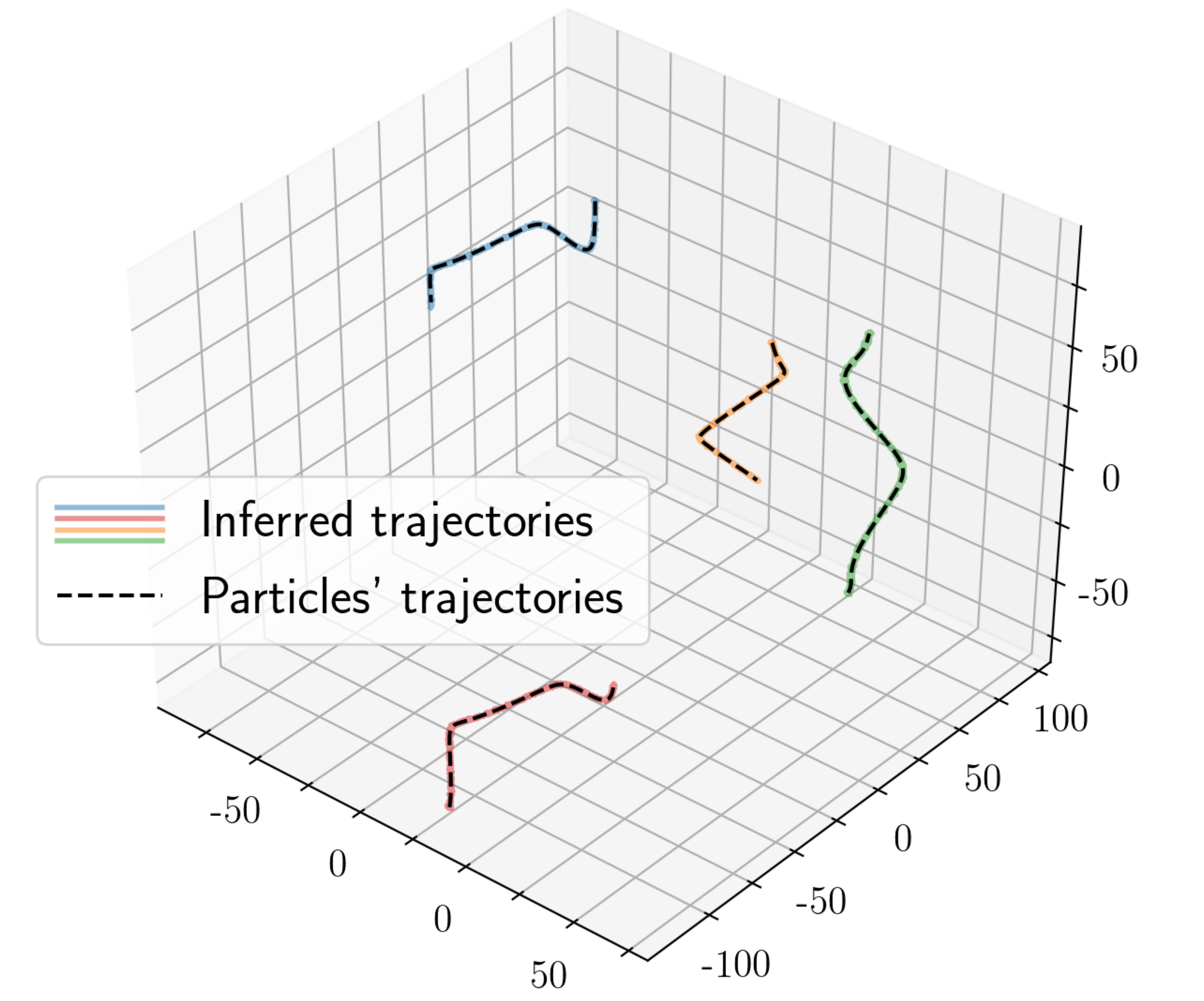}
        \caption{\label{fig:Simple_Track}Tracking of four particles. Trajectories inferred from synthetic LORs using the sequential PEPT-EM algorithm (colored lines) are compared with the actual trajectories (dashed lines). See section \ref{sec:Gate} for details of the simulated trajectories and LOR data.}
\end{figure}

The sequential tracking algorithm is summarized in Algorithm 2.
Figure \ref{fig:Simple_Track} illustrates its application by showing the reconstructed trajectories of four  particles moving in the field of view in simulated LOR data (details of the particle motion and LOR data generation are given in section \ref{sec:Gate}). The particles are correctly identified by the algorithm as soon as they enter the field of view and tracked accurately thereafter.
 

\begin{algorithm}[H]
    \caption{Sequential tracking algorithm}\label{algo:Tracking}
    \begin{algorithmic}[1]
        \State{Initialize $\{\bm{x}_{k,0},\sigma^2_{k,0},\rho_{k,0}\}_{1\leq k\leq K}$}
        \State{$n\gets0$}
        \Repeat\Comment{Loop over batches}
        \State{Update $\{\bm{x}_{k,n},\sigma^2_{k,n},\rho_{k,n}\}$ using EM algorithm over $\mathcal{L}_n$}
        \State{$\{\bm{x}_{k,n+1},\sigma^2_{k,n+1},\rho_{k,n+1}\}_{1\leq k\leq K}\gets\{\bm{x}_{k,n},\sigma^2_{k,n},\rho_{k,n}\}_{1\leq k\leq K}$}
        \State{$n\gets n+1$}
    \Until{completion}
    \end{algorithmic}
\end{algorithm}

\subsection{Higher-order tracking}

The sequential application of the PEPT-EM algorithm just described does not make use of the timing information of the LORs within each batch. Here, we propose a technique    which exploits this information. We parameterize the particle trajectories by the Taylor expansion
\begin{equation}\label{eq:Taylor_param}
    \bm{x}_k(t)=\sum_{m=0}^{M}\frac{(t-t_n)^m}{m!}\left.\frac{\mathrm{d}^m\bm{x}_k}{\mathrm{d}t^m}\right|_{t_n}
\end{equation}
about the batch's mean time $t_n=n\delta t$,
truncated at order $M$. It is useful to write \eqref{eq:Taylor_param} in the matrix form
\begin{equation}
    \bm{x}_k(t)=\mathbf{M}_n(t)\bm{X}_{n,k},
\end{equation}
where
\begin{equation}
    \mathbf{M}_n(t)\equiv
    \begin{bmatrix}
        \mathbb{I}_3\hspace{1em} (t-t_n)\mathbb{I}_3\hspace{1em} \dots\hspace{1em} {(t-t_n)^m} \mathbb{I}_3/m!
    \end{bmatrix}\quad
\end{equation}
and
\begin{equation}
    \\
    \quad\bm{X}_{k,n}=
    \begin{bmatrix}
    \left.\bm{x}_{k}\right|_{t_n}\\
    \left.{\mathrm{d}\bm{x}_k}/{\mathrm{d}t}\right|_{t_n}\\
    \dots\\
    \left.{\mathrm{d}^m\bm{x}_k}/{\mathrm{d}t^m}\right|_{t_n}
    \end{bmatrix}.
\end{equation}
Here, the $3\times 3 (M+1)$ matrix $\mathbf{M}_n(t)$ is given while the  $3(M+1)$-dimensional vector $\bm{X}_{k,n}$ has to be inferred.  

\begin{figure}
        \includegraphics[height=0.45\textwidth]{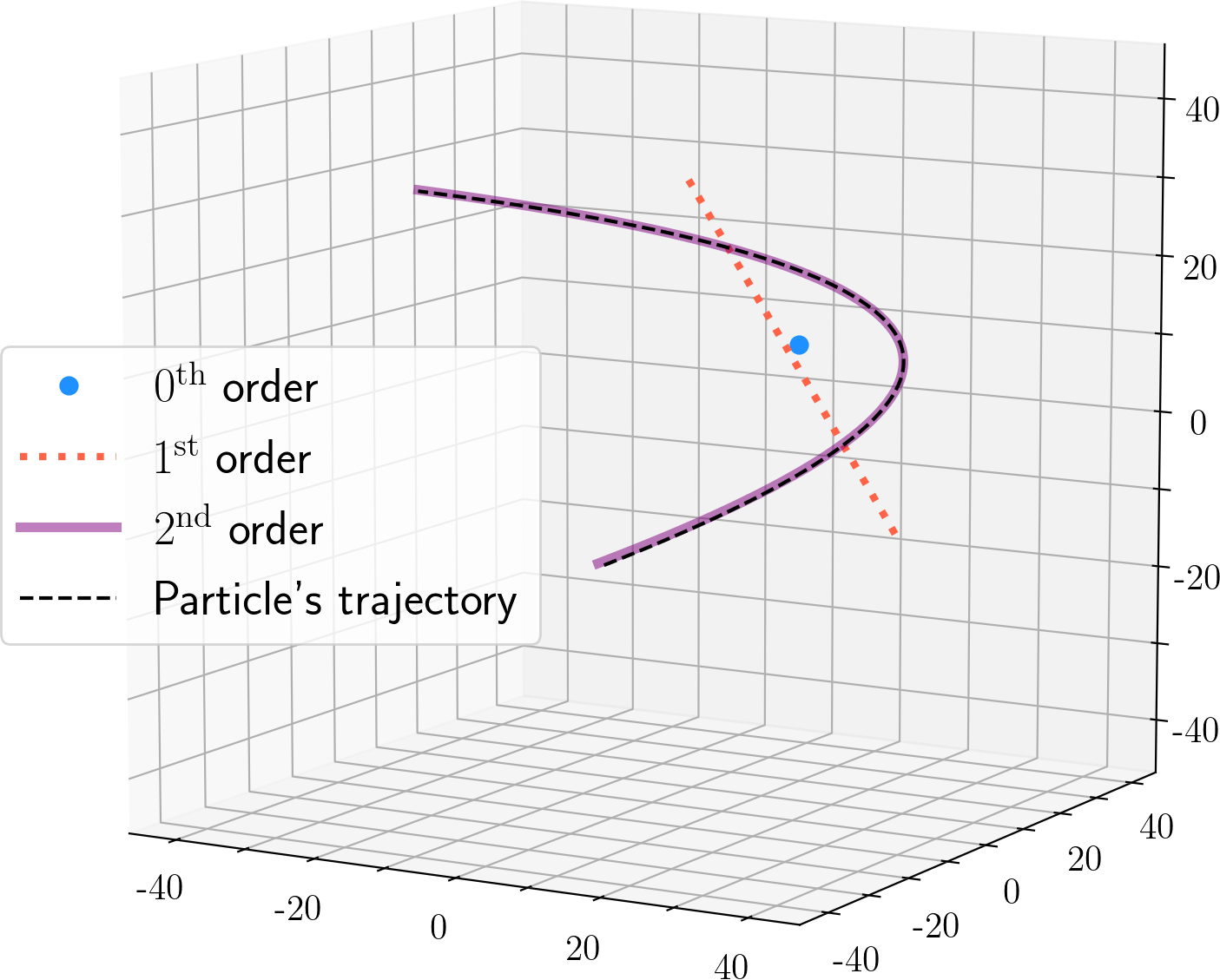} 
        \caption{\label{fig:High_Order_track}Higher-order tracking of a single particle from a single LOR batch. The exact (parabolic) trajectory  of the particle (black dashed  line) is compared with the approximations inferred using the PEPT-EM-0 (blue dot), PEPT-EM-1 (dotted red line)  and PEPT-EM-2 (solid magenta line) algorithms.}
\end{figure}

With the parameterization \eqref{eq:Taylor_param}, we can write the log-likelihood of  batch $n$ as
\begin{equation}\label{eq:LogLH_T}
   L\left(\{\bm{X}_{k,n},\sigma_{k,n}^2,\rho_{k,n}\}_{1\leq k\leq K}\right)\\=\sum_{\ell\in\mathcal{L}_n}\log\left(\rho_{0,n}\alpha+\sum_{k=1}^{K}\rho_{k,n}\sigma_{k,n}^{-2}\mathrm{e}^{-{D^2(\mathbf{M}_n(t_\ell)\bm{X}_{k,n},\ell)}/{2\sigma_{k,n}^2}}\right).
\end{equation}
This generalizes expression \eqref{eq:LogLH_alpha}, obtained assuming fixed positions and corresponding to $M=0$, to account for the motion of the particles within the time interval $\Delta t$ of each batch.
 We can extend the PEPT-EM algorithm to obtain an approximation of the parameters $\{\bm{X}_{k,n},\sigma_{k,n}^2,\rho_{k,n}\}_{1\leq k\leq K}$ maximizing this likelihood. A derivation paralleling that in section \ref{sec:Localizing} gives the latent weights as
\begin{equation}\label{eq:Laten_Weights_M}
    w_{\ell,k,n} =\frac{ \rho_{k,n}\sigma^{-2}_{k,n}\mathrm{e}^{-{D^2(\mathbf{M}_n(t_\ell)\bm{X}_{k,n},\ell)}/{2\sigma^2_{k,n}}}}{ \rho_{0,n}\alpha+\sum_{k'=1}^K\rho_{k',n}\sigma^{-2}_{k',n}\mathrm{e}^{-{D^2(\mathbf{M}_n(t_\ell)\bm{X}_{k',n},\ell)}/{2\sigma^2_{k',n}}}},
\end{equation}
the optimal positions as
\begin{equation}\label{eq:Centroid_weighted_M}         \bm{X}_{k,n} =   \mathbf{Centroid}_{k,n} \equiv \left(\sum_{\ell\in\mathcal{L}_n}w_{\ell,k,n}\mathbf{M}^\mathrm{T} _n(t_\ell)\mathbb{T}_{\ell}\mathbf{M}_n(t_\ell)\right)^{-1}\sum_{\ell\in\mathcal{L}_n}w_{\ell,k,n}\mathbf{M}_n^{\mathrm{T}}(t_\ell)\mathbb{T}_{\ell}\,\bm{y}_{\ell}
\end{equation}
with $\mathrm{T}$ denoting the matrix transpose,
and the optimal component weights as
 \begin{equation}\label{eq:R_M}
         \rho_{k,n} =   \mathrm{R}_k \equiv \frac{1}{N_n}\sum_{\ell\in L}w_{\ell,k,n},
        \end{equation}
with $N_n$ the number of LORs in batch $n$. In the following, we denote the order-$M$ algorithm consisting of the repeated application of the expectation step \eqref{eq:Laten_Weights_M} and maximization step \eqref{eq:Laten_Weights_M}--\eqref{eq:R_M} by ``PEPT-EM-$M$''. Figure \ref{fig:High_Order_track} illustrates the approximation of trajectories within batches used for PEPT-EM-$0$, -$1$ and -$2$. 


\begin{figure}
    \centering
    \includegraphics[width=\linewidth]{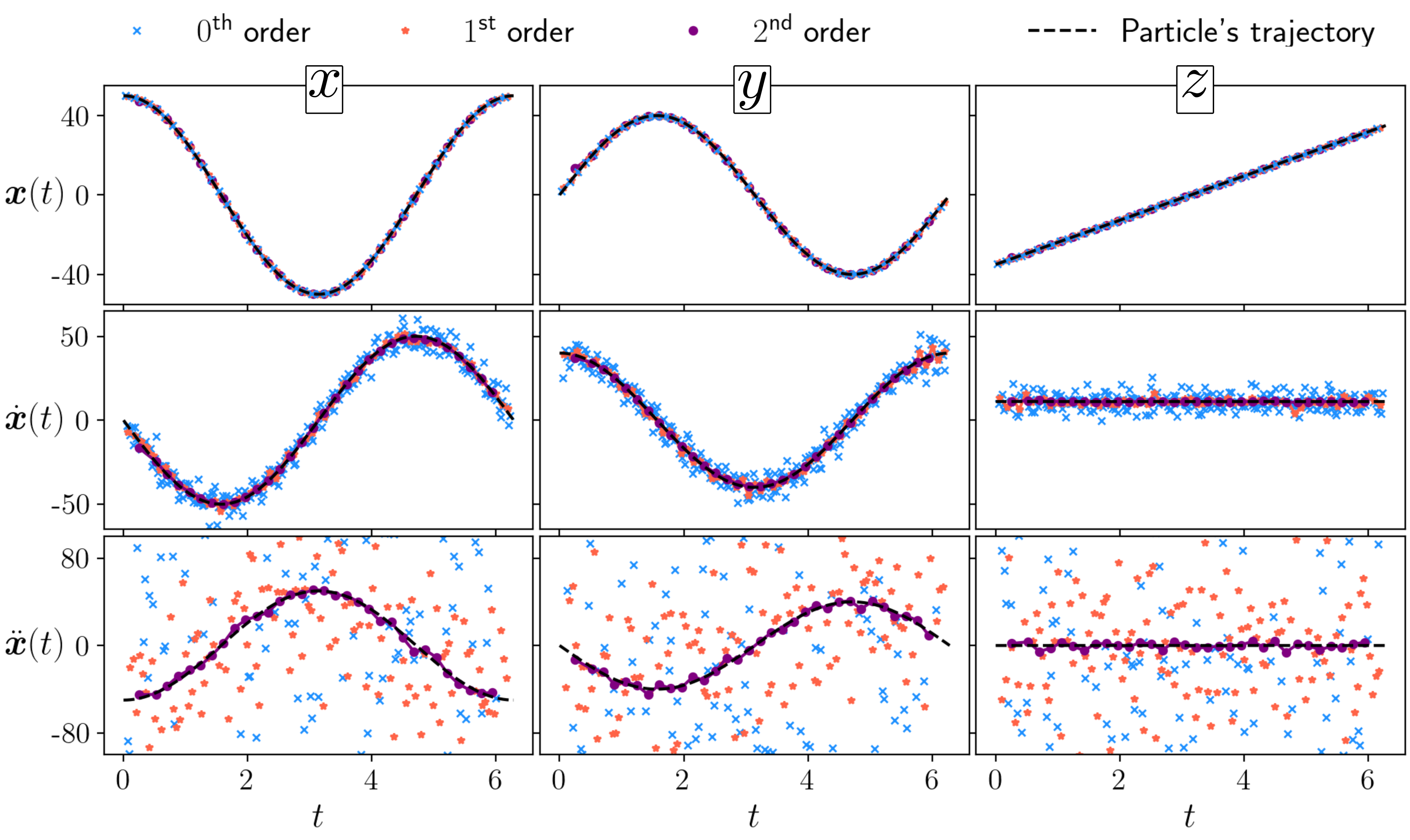}
    \caption{Velocity and acceleration estimation. Comparison of the position (top), velocity (middle) and acceleration (bottom) of a particle moving on a helical trajectory estimated using the PEPT-EM-0 (blue cross), PEPT-EM-1 (orange star) and PEPT-EM-2 algorithms (purple circle). A simple finite-difference approximation is used to estimate the velocity and acceleration when they are not directly provided by the algorithm. For  PEPT-EM-0 (resp. -1 and -2)  $10^5$ LORs are collected into $250$ (resp. $125$ and $32$) overlapping batches of $1000$ LORs (resp. $2000$ and $8000$). At every order, $K=2$ clusters are used (removing the one with the larger variance) and the parameter $\alpha$ is set to $0$. The dashed line shows the exact trajectory.}
    \label{fig:MultOrder}
\end{figure}

In sequential applications, the first guesses for the positions, velocities, etc.\ of a new batch can be obtained by extrapolating from the results of the previous batch. Figure \ref{fig:MultOrder} shows such a sequential application to a single (simulated) particle on an helical trajectory. The benefits of higher-order tracking are two-fold. First, the algorithm produces estimates for velocities and (for $M \ge 2$) accelerations directly, without the need for subsequent differentiation of the trajectory data. As figure \ref{fig:MultOrder} demonstrates, this leads to much improved estimates. Second, it also makes it possible to treat much bigger batch size, since this size is restricted by the assumption that the trajectory over the time interval $\Delta t$ is well approximated by a degree-$M$ polynomial rather than by the assumption of a fixed particle.
Overall, the accuracy of the prediction can be significantly increased by using the higher-order EM algorithm. 

\section{Application to PEPT data\label{sec:Application}}

In this section, we demonstrate the effectiveness of the PEPT-EM algorithm by applying it to several datasets. The first dataset is synthetic, generated by Monte Carlo simulations of particles moving in a simple prescribed  flow; the other datasets are generated using real data from the ADAC Forte Gamma Camera at the University of Birmingham \cite{Parker2002}. The Monte Carlo simulation was set up to emulate the ADAC Forte Gamma Camera, which we now introduce briefly.

\begin{figure}
    \centering
    \includegraphics[width=0.6\linewidth]{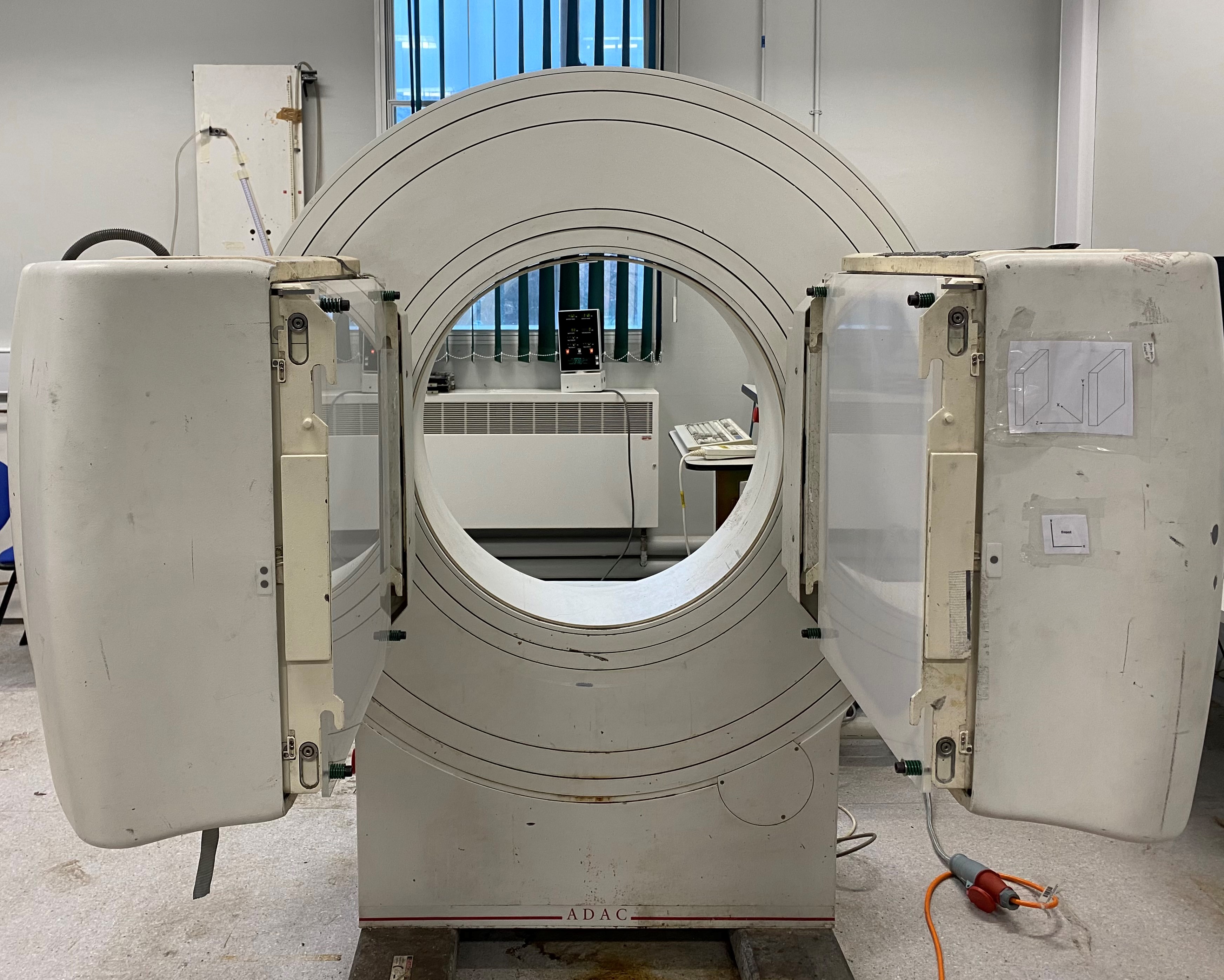}
    \caption{The ADAC Forte Gamma Camera at the University of Birmingham. Two scintillating detector screens are located on opposing sides of a cylindrical gantry.}
    \label{fig:ADAC_Camera}
\end{figure}

The ADAC Forte Gamma Camera, shown in figure \ref{fig:ADAC_Camera}, comprises two planes of scintillating detectors measuring approximately \(59\times 47 \text{ cm}^2\) with an adjustable separation between 25 to 80 cm. The detectors contain a 16 mm thick thallium-doped sodium iodide (Na(T)I) crystal optically coupled to an array of 55 photomultiplier tubes, with a software-based readout used to determine to location of each hit with a FWHM of approximately 6 mm. Coincident hits occurring within \(15 \, \text{ns}\) of each other are recorded in a listmode format.

\subsection{Simulated GATE data\label{sec:Gate}}
\begin{figure}
    \centering
    \includegraphics[width=0.6\linewidth]{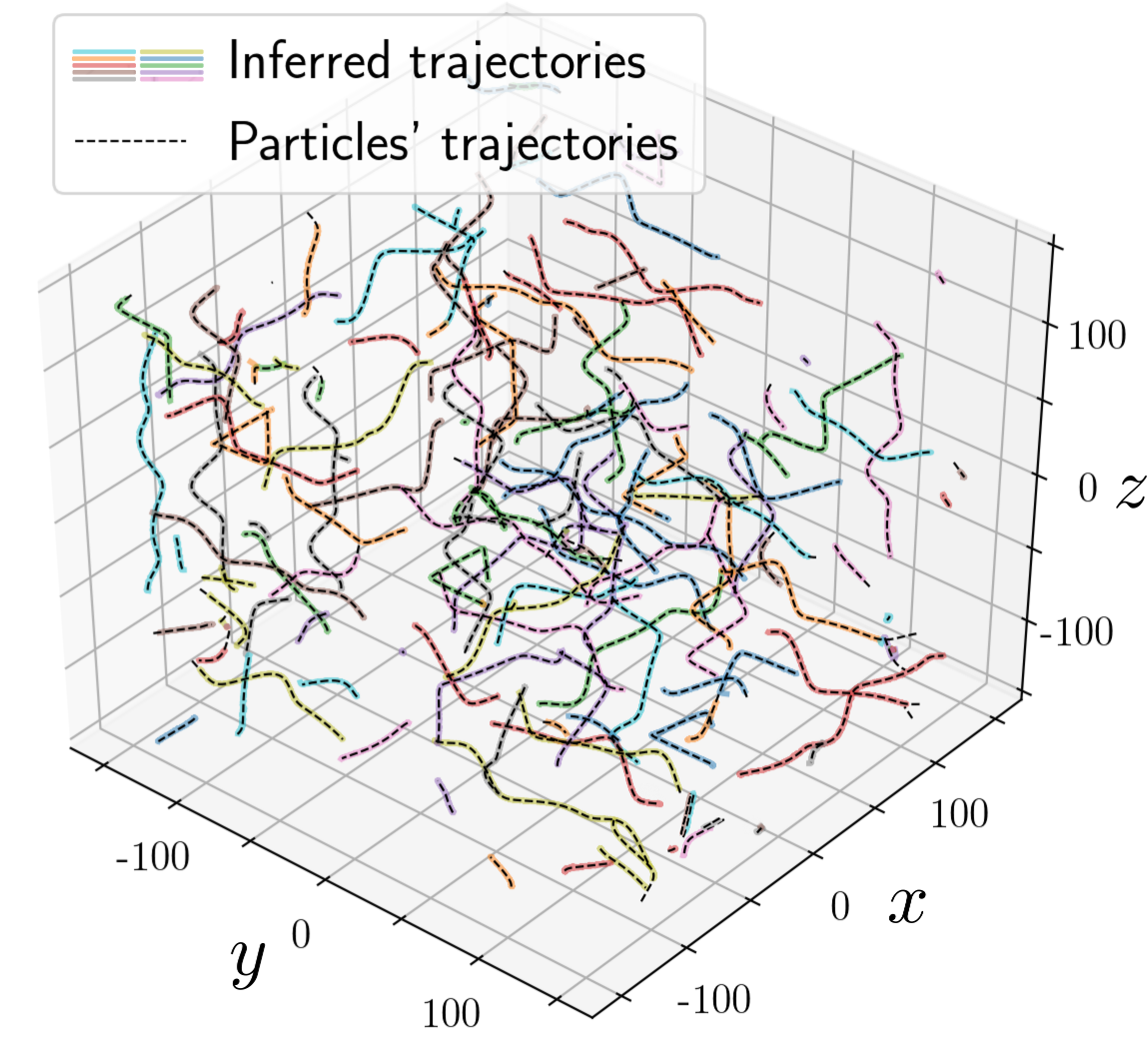}
    \caption{{Tracking of $80$ simulated particles.}  Comparison of particle trajectories inferred by the PEPT-EM-1 algorithm (colored) to the exact trajectories (dashed). The $10$ second-long particle trajectories are generated by an ABC-flow with parameters $A=B=C=0.01 \, \mathrm{m} \, \mathrm{s}^{-1}$ and wavelength $\lambda=50 \, \mathrm{mm}$, using random initial conditions and a periodic cubic domain $[-150,150]^3$ mm$^3$. The LORs are generated by a Monte Carlo model of the ADAC Forte Gamma Camera using the software package GATE\cite{herald2021}, assuming particles with an equal activity of $20 \, \mathrm{MBq}$. $K=160$ clusters are used and the LORs are collected into $90 \, \mathrm{ms}$-long successive frames with a $60 \, \mathrm{ms}$ overlap.}
    \label{fig:Gate_Tracking}
\end{figure}

In this first application of the PEPT-EM algorithm, we demonstrate the algorithm's ability to track large numbers of particles. We generated 80 trajectories of particles advected by the Arnold--Beltrami--Childress (ABC) flow \cite{Dombre1986} with random initial conditions. We use each trajectory as input for a Monte Carlo model of the ADAC Forte Gamma Camera at The University of Birmingham \cite{herald2021}. 
The Monte Carlo model was written using the GATE software package \cite{Jan2004}, used to replicate the characteristics of positron emission and scanners through the use of validated physics models. By providing the particle trajectories to the model, we obtain LORs matching closely those that would be seen in a real experiment, including scatter and some random coincidences. We then combined the LORs generated by all the particles into a single dataset. An advantage of this approach is that a large number of particles with well-defined positions and trajectories can be generated to devise a challenging test for the PEPT-EM algorithm. The approach avoids the difficulty that, in  a fully realistic set up where all the particles are simulated together, the near-simultaneous emissions by all the particles can lead to saturation of the detectors, a phenomenon which puts an upper limit on the number of particles that can be handled with current hardware. 

Figure \ref{fig:Gate_Tracking} shows the 80 exact particle trajectories and their reconstructed approximation obtained using PEPT-EM-1. The tracking is remarkably accurate, in spite of the large number of particles and their close proximity. The results strongly suggest that advanced reconstruction algorithms such as PEPT-EM can remove one of the barriers currently preventing the use of PEPT to track more than a handful of particles. 



\subsection{Ten rotating particles}

\begin{figure}
    \centering
     \begin{subfigure}[b]{0.4\textwidth}
         \centering
         \includegraphics[width=\textwidth]{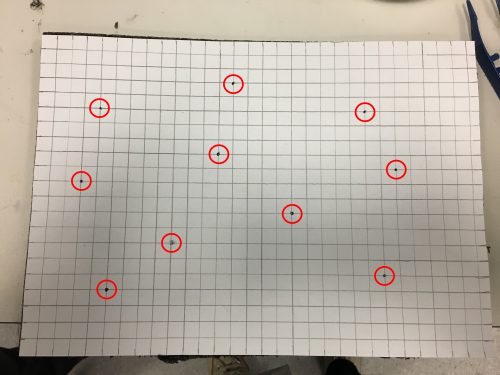}
         \caption{}
         \label{fig:locationsphoto}
     \end{subfigure}
      \begin{subfigure}[b]{0.4\textwidth}
         \centering
         \includegraphics[width=\textwidth]{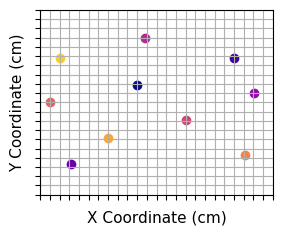}
         \caption{}
         \label{fig:locationsgraph}
     \end{subfigure}
    \caption{Photograph of the 10 particles attached to the grid with their locations highlighted (left), and  positions located using PEPT-EM-0 and static data acquisition (right).}
    \label{fig:StaticLocations}
\end{figure}

The following sections describe trajectory reconstruction with real experimental data. For the first experiment, we embedded 10 radiolabelled ion-exchange resin particles measuring approximately 0.3 mm in diameter  into a piece of foam, at well-spaced locations marked on a \(1 \, \textrm{cm}^2\) grid (see figure \ref{fig:locationsphoto}). Each tracer had an activity between 1 and 4 MBq. The foam was attached to a turntable to make the particles rotate. The turntable was placed in the field of view of the ADAC camera. A static data acquisition was performed first, followed by data acquisition with the turntable rotating at approximately 80 rpm.

\begin{figure}
    \centering
    \includegraphics[width=0.7\linewidth]{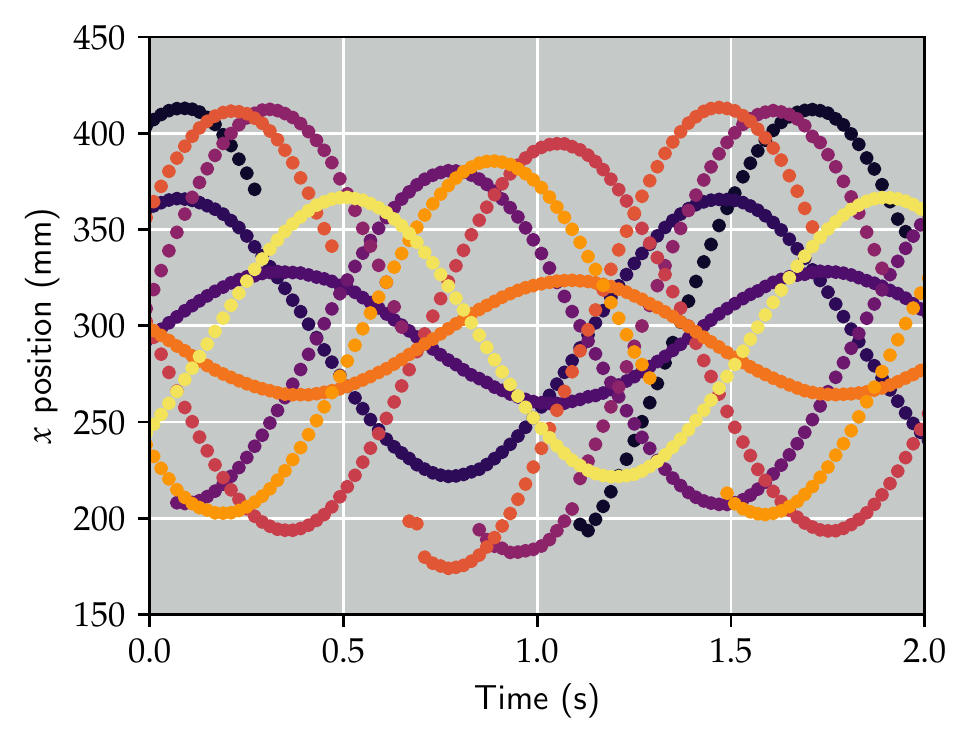}
    \caption{Trajectories tracked by PEPT-EM-2 in  experiment with 10 rotating particles. Discontinuities in the sinusoidal patterns indicate a temporary loss of tracking but overall the 10 trajectories are identified. Trajectories are colored according to the cluster label $k$.}
    \label{fig:Rotation_Tracking}
\end{figure}

\begin{figure}
    \centering
    \includegraphics[width=0.7\linewidth]{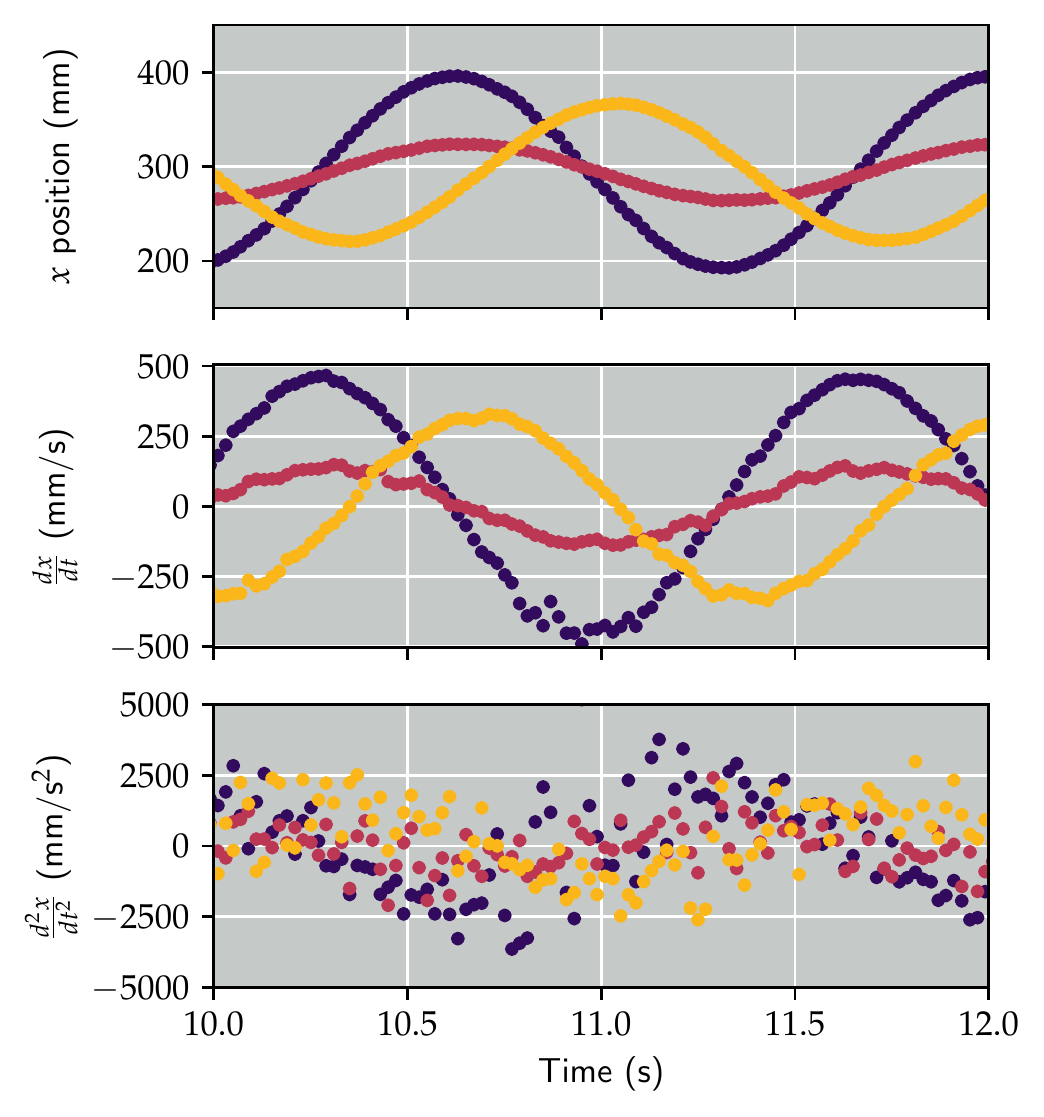}
    \caption{Tracked locations, velocities and accelerations of 3 selected particles from the 10 rotating-particle experiment in figure \ref{fig:Rotation_Tracking}.}
    \label{fig:Rotation_EM2}
\end{figure}

The LORs detected in each experiment are used to reconstruct the particle positions using PEPT-EM. The static set is analysed using PEPT-EM-0, as there is no velocity or acceleration. The positions found in this way are shown in figure \ref{fig:StaticLocations} and match closely the exact positions. The particle positions for the rotating experiment are tracked using PEPT-EM2, using 100 ms time batches with 80 ms overlap, such that we infer positions along the particle trajectories every 20 ms; \(K = 13\) clusters are used. The inferred positions are shown in figure \ref{fig:Rotation_Tracking}. 
The 10 particles are successfully tracked for most of the 2 s of the experiment, but there are breaks in the trajectories of 4 of the particles. These affect particles most distant from the axis of rotation, when they are close to the boundary of the field of view in the Y-axis and when they travel fastest. 
This highlights a limitation of the sequential approach in tracking high-speed particles, even though this is improved by PEPT-EM-2's use of velocity and acceleration as inferred parameters. In figure \ref{fig:Rotation_EM2}, we show the position, velocity and acceleration along 3 trajectories reconstructed by PEPT-EM-2. This shows that the algorithm provides a relatively smooth velocity but that the acceleration is significantly noisier. The noisy acceleration is in contrast with that of the single-particle experiment shown in figure \ref{fig:MultOrder}, reflecting the complexity introduced by multi-particle tracking.

We experimented using PEPT-EM-0 and PEPT-EM-1 to track the particles. We found  PEPT-EM-0 to be able to identify more particles simultaneously than PEPT-EM-1 and PEPT-EM-2, however with larger variances and errors. We also found that PEPT-EM-1 and PEPT-EM-2  were better at segregating ``inlier'' and ``outlier'' clusters, with a clearly defined gap in the variances. Overall, PEPT-EM-1 and PEPT-EM-2 performed at a similar level and significantly better than PEPT-EM-0.


\subsection{Pipe-flow experiment: merged trajectories} 


A small number of radioactive particles were placed into a flow loop to enable the tracking of liquid flow in a pipe. The number of particles is small enough that typically only 1 particle is in the field of view at any one time, allowing PEPT to be performed with single particle tracking using the Birmingham Method. We synthesize a new, more challenging data set by merging the LORs generated by 10 particles. We first choose 10 arbitrary trajectories from the single particle set; then, using their time stamps, we extract the LORs occurring as the particle passes the detector. For each trajectory, we then shift the time such that the first LOR of each set happens at \(t=0\). The LORs from each event are then  merged and sorted by time stamp to model a short window of time (5 seconds) where 10 particles cross the field of view simultaneously.

\begin{figure} 
    \centering
         \includegraphics[width=0.4\textwidth]{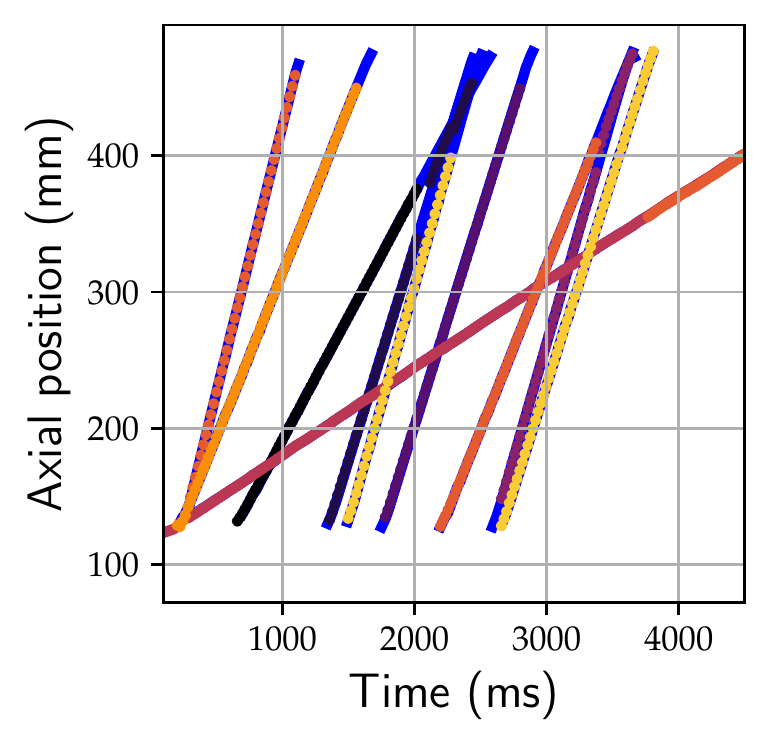}
      \hfill
         \includegraphics[width=0.55\textwidth]{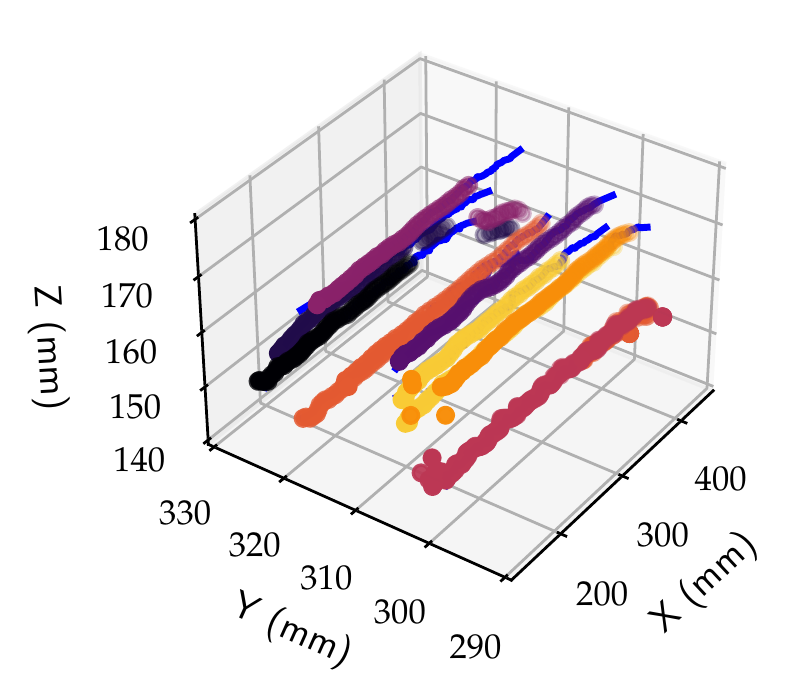}
    \caption{Trajectories of 10 particles in the pipe-flow experiment: the blue curves show the trajectories of particles tracked individually using the Birmingham algorithm; the other curves show the trajectories reconstructed by the PEPT-EM1 algorithm with data merging the LORs generated by the 10 particles. The figures show the along-pipe coordinate $x$ as a function of time (left) and  three-dimensional trajectories (right).}
     \label{fig:pipe_tracks}
\end{figure}

This merged data set is then tracked using PEPT-EM1, using \(K = 10\) clusters, LOR batches that are 50 ms with an overlap of 30 ms. We compare each reconstructed trajectory using multiple particle tracking from PEPT-EM1 with those from single particle tracking using the Birmingham Method. In a single instance, a maximum of 5 particles are in the field of view in our merged dataset. The  trajectories are shown in figure \ref{fig:pipe_tracks}. They are separated simply using the cluster label attributed by PEPT-EM1. Note that the very short trajectories seen in the 3D plot in figure \ref{fig:pipe_tracks} are the tails of trajectories that weren't seen in the Birmingham algorithm when selecting the LORs.

Some simple analysis was performed to compare the trajectories produced by PEPT-EM and the Birmingham Algorithm. The average location of each particle was recorded in 50 ms bins across the length of each trajectory. This was done for both PEPT-EM1 and the Birmingham Algorithm so that the two trajectories could be directly compared. The RMS difference in position averaged over all particles is 1.7 mm in the axial (along-pipe) direction, 0.3 mm in the vertical direction, and 0.8 mm in the horizontal direction. Since the typical uncertainty on a PEPT experiment using the ADAC camera is of the order of 1 mm, we consider the results using multiparticle tracking with PEPT-EM1 on this dataset to be comparable with those produced using the Birmingham method on arguably cleaner data.
We note that in figure 11 some trajectories are not completely reconstructed when simply identifying them by their cluster label. This could be improved in post-processing by using more advanced trajectory separation.

\subsection{Pipe-flow experiment: flow-profile reconstruction}

The final data set comprises real experimental data used to measure the velocity profile of liquid flow in a pipe. 10 quasi-neutrally buoyant ion exchange resin particles measuring approximately 0.3 mm in diameter, with initial activities between 20-30 MBq were used to track the flow in a set up similar to that of the previous section. The data was acquired over a time period of 90 min, with typically no more than one tracer in the camera field-of-view at any one time. The raw LOR data was sliced into 5 minute chunks and then merged into a single data set of length 5 minutes in order to model multiparticle tracking. In this case, the data were produced not just to show that PEPT-EM can track several particles simultaneously, but that it can be used to characterize the bulk characteristics of a flow in a single short experiment.

\begin{figure}
\centering
    \includegraphics[width=0.7\linewidth]{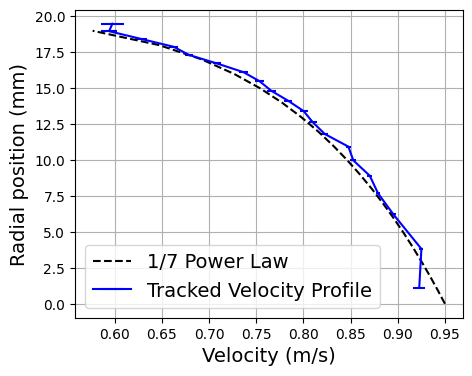}
    \caption{Velocity profile constructed using PEPT-EM1 from LOR generated by 10 particles in a pipe flow. A 1/7th power law profile is overlaid.}
    \label{fig:velocity_profile}
\end{figure}

Particles trajectories are reconstructed using PEPT-EM1, with a batch time interval of 50 ms, an overlap of 30 ms, and \(K = 10\) clusters. The algorithm provides an estimate for the position, velocity and variance of each cluster for each batch. Clusters with sizes exceeding 10 mm are considered as outliers. The inferred along-pipe velocities are then binned as a function of the radial coordinate of the pipe to obtain the mean-flow profile shown in figure \ref{fig:velocity_profile}. The profile is compared with the 1/7th power law velocity profile that is typically used to describe velocity profiles in turbulent pipe flows\cite{DeChant2005}, with good agreement. Our experiments demonstrates the potential for PEPT-EM1 to vastly reduce the experimental time required needed to obtain a velocity profile, as well as a reduction in the calculation needed to produce such a profile.

\section{Conclusion \label{sec:Conclusion}}

In this paper, we introduce and assess a novel algorithm, PEPT-EM, for tracking multiple particles in PEPT. The algorithm relies on the maximization of a likelihood associated with a simple model of positron annihilation and photon emission, using the standard expectation--maximization (EM) method to estimate the maximising parameters and hence the particle positions. The algorithm is best applied sequentially, using a sequence of batches of LORs to reconstruct successive particle positions. Using the outcome of one batch as a first guess for the maximization for the next batch greatly accelerates the convergence of the EM algorithm. Moreover, it naturally leads to the cluster labels $k$ remaining attached to the same particle across batches, so entire trajectories are reconstructed with minimal post-processing.  
Higher-order tracking, which infers derivatives of the positions, improves this further and yields smooth approximations for the velocities and possibly accelerations along trajectories. 
We note that PEPT-EM also provides an estimate for the (relative) activity of each particle. This provides a way of distinguishing particles, potentially useful in case of collision as recently proposed in Ref.\ \citenum{Nicusan2020}.


Multi-particle tracking is still a relatively novel development in the world of PEPT. The difficulty of identifying individual particles, some in close proximity, with accuracy and reliability have prevented the technique being used routinely. We emphasize the strong benefits of multi-particle tracking: not only does it greatly speed up the data acquisition process, it also makes it possible to examine spatial correlations in fluid flows, as is essential to understand turbulent phenomena or particle--particle interactions. The PEPT-EM algorithm is one of several recently proposed algorithms\cite{wiggins2016,wiggins2017,Blakemore2019,Nicusan2020} which can transform the range of applicability of PEPT by enabling the tracking of large numbers of particles. A comparison of several algorithms including PEPT-EM, assessing their performances on a series of benchmark tests, is currently underway. It will be the subject of a future review paper. We hope that progress in trajectory-reconstruction algorithms will stimulate new improvements in detection hardware, in particular increasing the maximum data acquisition rate which, at present, remains a limiting factor in tracking large numbers of particles.

We conclude by emphasizing the Bayesian grounding of the PEPT-EM algorithm, which is based on a (forward) model for the generation of LORs by particles with given activities and given locations. This provides a framework to incorporate prior information, e.g. bounds on the radioactivity of the particles, and to go beyond the simple maximization to quantify the uncertainty of the reconstructed trajectories. Work in this direction is also underway. 




\begin{acknowledgments}
This research was supported  was supported by EPSRC Programme Grant EP/R045046/1: Probing Multiscale Complex Multiphase Flows with Positrons for Engineering and Biomedical Applications (PI: Prof. M. Barigou, University of Birmingham). The authors would like to thank Matthew Herald for running the Monte Carlo simulations of the ADAC camera, and Mostafa Barigou, Ananda Jadhav, Kun Li, Chiya Savari, Hamzah Sheikh and Zhuangjian Yang for their work on designing and setting up the pipe flow rig experiments and recording the raw PEPT data.
\end{acknowledgments}

\section{Data Availability Statement}
The data that support the findings of this study are available from the corresponding author upon reasonable request, pending permission from the PI and any further use of data must explicitly state the source.

\bibliography{References}
\end{document}